\documentclass[twocolumn,10pt,cleanfoot]{asme2e}
\input{Nomenclature.tex}
\usepackage{bm}
\usepackage{graphicx}
\usepackage{amsmath}
\usepackage{hyperref}
\usepackage{amssymb}
\usepackage{mathtools}
\newcommand{\numberset}[1]{\mathbb{#1}} 
\newcommand{\reals}{\numberset{R}}

\usepackage{lineno}
\usepackage{subcaption}
\usepackage{adjustbox}
\usepackage{lipsum}
\usepackage{tikz}
\usetikzlibrary{shapes,shapes.misc,decorations.markings,chains,positioning,arrows,calc}
\usepackage{array,multirow}
\usepackage{pgfplots}
\usepackage{gensymb}
\usepackage{amsbsy}
\pgfplotsset{compat=1.12}
\usepackage{subcaption}
\usepackage{dashbox}
\usepackage{filecontents}

\usepackage{array}
\newcolumntype{L}[1]{>{\raggedright\let\newline\\\arraybackslash\hspace{0pt}}m{#1}}
\newcolumntype{C}[1]{>{\centering\let\newline\\\arraybackslash\hspace{0pt}}m{#1}}
\newcolumntype{R}[1]{>{\raggedleft\let\newline\\\arraybackslash\hspace{0pt}}m{#1}}

\def\b{\mathbf}

\newcommand{\bs}{\boldsymbol}

\confshortname{OMAE 2019}
\conffullname{the ASME 38th International Conference on Ocean, Offshore \& \\ Arctic Engineering}

\confdate{9-14}
\confmonth{June}
\confyear{2019}
\confcity{Glasgow}
\confcountry{Scotland, UK}

\papernum{Under consideration to be published in OMAE 2019}

\title{DRAFT: A hybrid data-driven deep learning technique for fluid-structure interaction}

\begin{document}

\author{T. P. Miyanawala
    \thanks{Address all correspondence to this author.}
	\affiliation{Department of Mechanical Engineering\\
	National University of Singapore\\
	Singapore 117575\\
    Email: e0010779@u.nus.edu\\}}

 \author{Rajeev K. Jaiman
    \affiliation{Department of Mechanical Engineering\\
	University of British Columbia\\
	Vancouver, Canada\\
    Email: rjaiman@mech.ubc.ca}}

\maketitle 

\begin{abstract}
This paper is concerned with the development of a hybrid data-driven technique for unsteady fluid-structure interaction systems. The proposed data-driven technique combines the deep learning framework with a projection-based low-order modeling. 
While the deep learning provides low-dimensional approximations from datasets arising from black-box solvers, the projection-based model constructs the  low-dimensional approximations by projecting the original high-dimensional model onto a low-dimensional subspace.
Of particular interest of this paper is to predict the long time series of unsteady flow fields of a freely vibrating bluff-body subjected to wake-body synchronization. 
%
We consider convolutional neural networks (CNN) for the learning dynamics of wake-body interaction, which assemble layers of linear convolutions with nonlinear activations to automatically extract the low-dimensional flow features.
Using the high-fidelity time series data from the stabilized finite element Navier-Stokes solver, we first project the dataset to a low-dimensional subspace by proper orthogonal decomposition (POD) technique. 
The time-dependent coefficients of the POD subspace are mapped to the flow field via a CNN with nonlinear rectification, and the CNN is iteratively trained using the stochastic gradient descent method to predict the POD time coefficient when a new flow field is fed to it. 
The time-averaged flow field, the POD basis vectors, and the trained CNN are used to predict the long time series of the flow fields and the flow predictions are quantitatively assessed with the full-order (high-dimensional) simulation data.
The proposed POD-CNN model based on the data-driven approximation has a remarkable accuracy in the entire fluid domain including the highly nonlinear near wake region behind a freely vibrating bluff body.  
\end{abstract}

\begin{nomenclature}[$C_Dp$, $C_Lp$]
\entry{$\b{A}_i$}{POD time coefficients at time $t_i$}
\entry{$C$}{Translational damping coefficient}
\entry{$C_{Dp},C_{Lp}$}{Pressure-induced drag and lift coefficients}
\entry{$D$}{Characteristic length}
\entry{$\b{E}_i$}{CNN prediction error}
\entry{$f_n$}{Natural frequency of translational vibrations}
\entry{$\gamma$}{Learning rate of CNN}
\entry{$K$}{Translational stiffness coefficients}
\entry{$\b{K}$}{Convolutional kernel matrix}
\entry{$m$}{Mesh count}
\entry{$M$}{Mass of the bluff body}
\entry{$m^*$}{Mass ratio}
\entry{$m^f$}{Mass of the fluid displaced}
\entry{$\mu^f$}{Dynamic viscosity of the fluid}
\entry{$n$}{Number of snapshots}
\entry{$\b{n}$}{Unit normal vector}
\entry{$P_i$}{Normalized pressure of the $i^{th}$ node of the mesh}
\entry{$\psi$}{Momentum of CNN}
\entry{$\rho^f$}{Density of the fluid}
\entry{$Re$}{Reynolds number}
\entry{$\bs{\sigma_p}$}{Pressure contributed stress component}
\entry{$U_{\infty}$}{Fluid stream velocity}
\entry{$U_r$}{Reduced velocity for translational motion}
\entry{$\bs{\mathcal{V}}$}{POD mode matrix}
\entry{$\b{w}$}{Weight matrix of CNN}
\entry{$\bs{\mathcal{W}}$}{Kernels and weights of CNN ($\b{K}$ and $\b{w}$)}
\entry{$\b{Y}$}{Flow field dataset}
\entry{$\b{\overline{y}}$}{Mean flow field}
\entry{$\b{y}_i$}{Flow field snapshot at time $t_i$}
\entry{$\b{z}_i^c, \b{z}_i^r$}{Outputs of convolution and rectification layers}
\entry{$\zeta$}{Translational damping ratio}
\end{nomenclature}

\section{Introduction}
Unsteady flows involving fluid-structure interactions (FSI) are ubiquitous in  
marine and offshore engineering.
Even a simple configuration of a coupled FSI system can exhibit complex spatial-temporal dynamics and synchronization as functions of physical parameters and geometric variations.
Development of an efficient and a general purpose FSI model for the prediction and analysis of such unsteady dynamical behaviour is one of the long-standing problems. In particular, a reliable prediction of the underlying dynamics due to the vortex shedding and the wake-body synchronization poses a serious challenge to the current state-of-the-art analytical tools. 
Moreover, complex spatial-temporal characteristics are strongly dependent on the underlying physical parameters and geometric variations in such a coupled dynamical system of wake-body interaction.

While the current full-order modeling (i.e., solution of partial differential equations) techniques provide a high-fidelity data, they are time-consuming and computationally expensive for the prediction and analysis of long-term dynamics.
Furthermore, once generated, these high-fidelity data are rarely utilized to increase the efficiency of the subsequent flow field extraction and prediction. 
Herein, we are interested in the development of a general data-driven low-dimensional model that can learn the dynamical system well enough to efficiently forecast the time series of flow fields using the high-dimensional simulation data. 
For this purpose, we combine the projection-based model reduction with the deep learning, whereas the projection-based low-order approximation provides physically interpretable flow fields and the deep learning enables efficient prediction of unsteady flow fields.
During the development of this machine learning model for dynamical predictions, we surmise that the dominant flow features lie on an embedded nonlinear low-dimensional manifold within the high-dimensional space.

In the field of numerical analysis, the dimensionality reduction based on proper orthogonal decomposition (POD) provides a  low-dimensional data-driven approximation for analyzing dynamical systems \cite{rowley2017model}.
A high-fidelity dataset can be projected to a low-dimensional subspace via Galerkin method whereas the POD-Galerkin gives time-invariant spatial modes/basis related to significant physical features such as vortex street, shear layer and near-wake bubble \cite{miyanawala2018decomposition}. This low-dimensional basis can be used to reconstruct the already available field and extract the significant flow features which influence the underlying dynamics. 
%

In the field of computer science, one of the recent approaches for handling high-dimensional data is to learn the hidden representation automatically by supervision with the output via neural networks.
Such machine learning techniques, especially the biologically-inspired (neuron-based) learning, have a predictive capability for the time series estimation. In particular, convolution based neural networks with nonlinear rectification known as convolutional neural networks (CNN) have some attractive properties such as local connectivity, nonlinear embedded mapping and parameter sharing. In a recent work of \cite{miyanawala2017efficient}, the authors have demonstrated the ability of CNN for unsteady force prediction along with its physical analogy and justifications for the deterministic analysis of flow dynamics using the Navier-Stokes equations.
It is natural to integrate the physically interpretable POD-based low-order model  with a data-driven machine learning. 
The present study revolves around two fundamental questions: 
(i) can we utilize POD approximation to learn low-dimensional features for predicting unsteady laminar wake fields?
(ii) can we generalize combined POD-CNN for the simulation of wake-body synchronization while assuming the interaction of dominant low-dimensional features with a vibrating bluff body? 

In this paper, we combine the POD and CNN techniques in a way to take the advantage of the optimal low-dimensional representation given by POD and the multi-scale feature extraction provided by the CNN.
More specifically, we develop an efficient CNN model which is trained to learn the nonlinear functional relationship between the present flow field (input) and time-varying POD coefficients of the future time step (output). During the training phase, the CNN is fed with a set of known input-output combinations obtained by the full-order model (FOM). Using the trained CNN, we can feed a known flow field and predict the unknown time coefficients of the next time step. 
%
We apply the proposed POD-CNN technique to wake-body interaction systems which have statistically stationary dynamical behavior.
We consider a canonical problem of a laminar unsteady wake interacting with an elastically-mounted square cylinder because it offers distinct flow features that are linearly independent and can be learned to predict future flow fields.
We assess the predictions of the long time series of flow fields, namely global accuracy of the full domain, local resolution of the non-linear wake regions, the accuracy of the derived statistics such as the fluid forces on the bluff body. 
%
Predictions of unsteady wake-body interaction via low-dimensional learning model have a broad range of engineering applications with regard to design optimization, parameter space exploration and the real-time control and monitoring.

The outline of the rest of the article is as follows. The hybrid deep-learning model and the convolution based neural network are detailed in the next section. Section 3 presents the results and discussion together with the problem set-up, the hyperparameter tuning of the deep learning network and the performance of the hybrid deep learning model for fluid flow predictions. Finally, the major conclusions of this work are reported in the last section. 

\section{Low-dimensional deep learning model}
Herein, we present a low-dimensional learning model which is trained using the full-order simulation data to reconstruct the unknown future flow fields efficiently without solving the governing differential equations. While we utilize POD for the low-dimensional approximation, we consider convolutional neural networks as the learning technique, wherein we do not explicitly map adjacent flow fields.
To train a CNN explicitly for such high-dimensional many-to-many mapping, there is a need for significantly many convolution layers and/or kernels. When the number of convolution layers is increased and/or larger kernels are used, a weight in the final layer of the CNN represents a very large size problem. In this brute-force manner, the CNN loses the local connectivity which is essential to capture smaller length scale features present in the unsteady separated flows. Moreover, it makes the CNN extremely less efficient to train and execute for useful predictions.
We further assume that the solution space of the unsteady wake system attracts a low-dimensional subspace, which allows building a set of POD basis vectors from the
high-dimensional space. We map the time-dependent fluctuations of this low-dimensional representation with the known flow field data to reconstruct and predict the unknown flow fields via machine learning. By combining the dimensional reduction of the POD and the locally-connected learning process of CNN, we next describe our novel prediction technique for wake-body interaction systems.

\subsection{Hybrid POD-CNN procedure}
The prediction technique requires a high-fidelity series of snapshots of the flow field obtained by full-order simulations, experiments or field measurements.
Let $\b{Y}=\{\b{y}_1 \: \b{y}_2 \: ... \: \b{y}_n \} \in \reals^{m\times n}$ be the flow field data set where $\b{y}_i \in\reals^{m}$ is the flow field snapshot at time $t_i$ and $n$ is the number of snapshots. $m \gg n$ is the number of data points, for example, number of probes in an experiment or field measurements or the number of mesh nodes in a numerical simulation.
The target is to reconstruct and predict the future flow fields: $\b{y}_{n+1}, \b{y}_{n+2}, ... $ using the data set $\b{Y}$.
The proposed POD-CNN technique can be divided into four key steps, which are as follows:
\\

\noindent
\emph{Step 1: Generate the proper orthogonal decomposition (POD) basis for dataset $\b{Y}$}

Using the $n$ snapshots we can determine the mean field ($\overline{\b{y}}\in \reals^m$), the POD modes ($\bs{\mathcal{V}}\in \reals^{m\times k}$) and the time dependent POD coefficients, such that
\begin{equation}
\b{y}_i \approx \overline{\b{y}} +\bs{\mathcal{V}}\b{A}_i, \qquad i=1,2,...,n
\end{equation}
where $\b{A}_i=[a_{i1} \: a_{i2} \: ... \: a_{ik}] \in \reals^k$ are the time coefficients of the first $k$ significant modes for the time $t_i$. The vectors $\{\b{A}_1, \b{A}_2, ..., \b{A}_n\}$ can be determined using the POD reconstruction technique based on linear or nonlinear projections. Note that $k\leq n \ll m$ and $k$ can be estimated via the mode energy distribution. This decomposition reduces the order of the flow field from $O(m)$ to $O(k)$.
Now the problem of predicting $\b{y}_{n+1}\in\reals^{m}$ simply reduces to the determination of the vector $\b{A}_{n+1}\in\reals^{k}$.
\\

\noindent
\emph{Step 2: Train the convolutional neural network}

Training consists of determining a function that allows to estimate a finite time series of measurements. We employ the convolutional neural network described Section in \ref{CNN} as the machine learning method. The CNN is employed between the flow field data set $\b{y}_{i-1}$ and the POD time coefficients $\b{A}_i$. We can construct a mapping using the following functional relationships:
\begin{equation}
\b{A}_i = f(\b{y}_i), \qquad \b{y}_i = g(\b{y}_{i-1}), \qquad \b{A}_i = f \circ g (\b{y}_{i-1}),
\end{equation}
where $f$ is the nonlinear function relating the time coefficients and the flow field data. For example, in the linear POD reconstruction, $f$ is the inner product $\langle \b{y}_i - \overline{\b{y}}, \bs{\mathcal{V}}\rangle$ and $g$ is the unknown mapping between the flow fields of adjacent time steps.
Set the convolutional neural network such that the data set $\{\b{y}_1,\b{y}_2,...,\b{y}_{n-1}\}$ is mapped to the POD coefficient set $\{\b{A}_2, \b{A}_3,...,\b{A}_n\}$ so that $f \circ g$ map is transformed to
\begin{equation}
\b{A}_i = F(\b{y}_{i-1},\b{K},\b{w}),
\end{equation}
where $\b{K}$ and $\b{w}$ are the time-independent trained kernels and weights and $F$ denotes the CNN operation. The CNN is initialized with guessed convolution kernels and weights. The network predicts the outputs and compares them with the real output and iteratively adjusts the kernels and weights. Once trained, $\b{K}$ and $\b{w}$ form an offline network which outputs the POD coefficients of the next time step when fed with the flow field matrix.
\\

\noindent
\emph{Step 3: Prediction of $\b{A}_{n+1}$}

In Step 2, it is worth noting that the known flow field of the last time step $\b{y}_n$ is not used for the CNN training. By feeding it to the trained CNN, we can obtain the unknown POD time coefficients at $t_{n+1}$: $\b{A}_{n+1}$. Using $\b{A}_{n+1}$, we can reconstruct the predicted flow field at $t_{n+1}$ (${\b{y}}_{n+1}$) as follows:
\begin{equation}
{\b{y}}_{n+1} \approx \overline{\b{y}} +\bs{\mathcal{V}}\b{A}_{n+1}.
\end{equation}
\\
\noindent
\emph{Step 4: Loop through the offline database and predict all other steps}

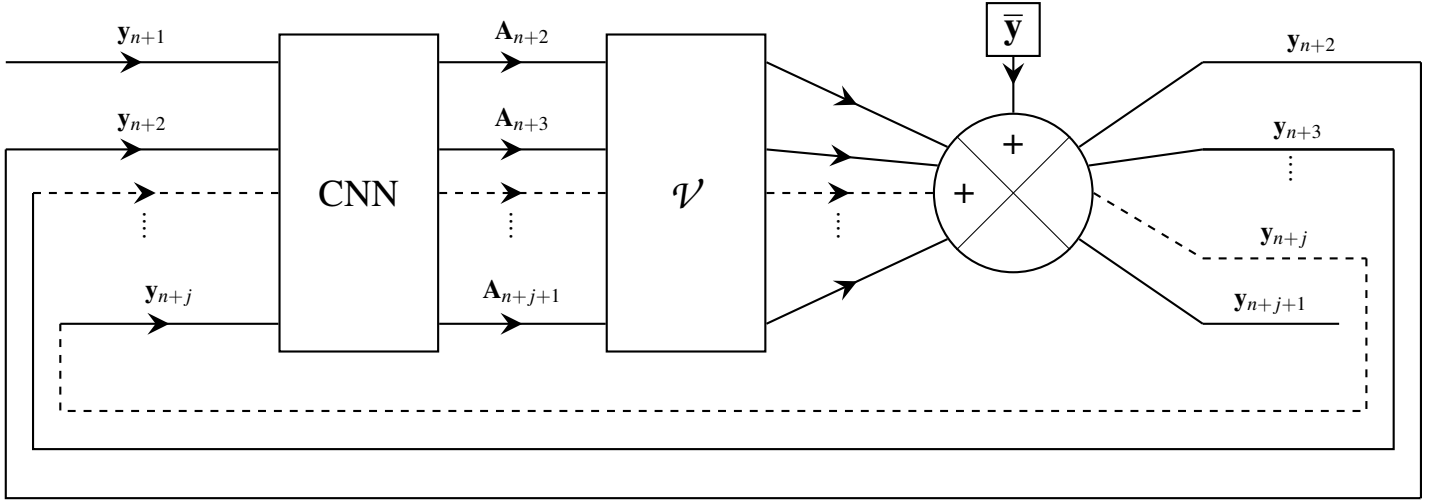
\begin{figure*}
\centering
   	\begin{tikzpicture}[decoration={markings,mark=at position 0.5 with {\arrow[scale=2]{>}}},scale=1.45]
   	\node[draw, thick, minimum width=60pt, minimum height=120pt, align=center] (CNN) at (0,0) {\Large{CNN}};
   	\draw [>=stealth,-, postaction={decorate}, thick, midway, above] ($(CNN.west)+(-2.5,1.2)$)-- node [midway,above,yshift=0.1cm] {$\b{y}_{n+1}$} ($(CNN.west)+(0,1.2)$) ;
    \draw [>=stealth,-, postaction={decorate}, thick, midway, above] ($(CNN.west)+(-2.5,0.4)$)-- node [midway,above,yshift=0.1cm] {$\b{y}_{n+2}$} ($(CNN.west)+(0,0.4)$) ;
    \draw [>=stealth,-, postaction={decorate}, thick, dashed, midway, above] ($(CNN.west)+(-2.25,0)$)-- node [midway,below left,yshift=-0.1cm, xshift=-0.3cm, rotate = 90] {$....$} ($(CNN.west)+(0,0)$) ;
    \draw [>=stealth,-, postaction={decorate}, thick, midway, above] ($(CNN.west)+(-2,-1.2)$)-- node [midway,above,yshift=0.1cm] {$\b{y}_{n+j}$} ($(CNN.west)+(0,-1.2)$) ;
    
    \node[draw, thick, minimum width=60pt, minimum height=120pt, align=center] (POD) at (3,0) {\Large{$\bs{\mathcal{V}}$}};
    \draw [>=stealth,-, postaction={decorate}, thick, midway, above] ($(CNN.east)+(0,1.2)$)-- node [midway,above,yshift=0.1cm] {$\b{A}_{n+2}$} ($(POD.west)+(0,1.2)$) ;
    \draw [>=stealth,-, postaction={decorate}, thick, midway, above] ($(CNN.east)+(0,0.4)$)-- node [midway,above,yshift=0.1cm] {$\b{A}_{n+3}$} ($(POD.west)+(0,0.4)$) ;
    \draw [>=stealth,-, postaction={decorate}, thick, dashed, midway, above] ($(CNN.east)+(0,0)$)-- node [midway,below left,yshift=-0.1cm, xshift=-0.3cm, rotate = 90] {$....$} ($(POD.west)+(0,0)$) ;
    \draw [>=stealth,-, postaction={decorate}, thick, midway, above] ($(CNN.east)+(0,-1.2)$)-- node [midway,above,yshift=0.1cm] {$\b{A}_{n+j+1}$} ($(POD.west)+(0,-1.2)$) ;
    
    \node[draw, thick, circle, minimum width=60pt, minimum height=50pt, align=left] (ADD) at (6,0){};
    \node[minimum width=20pt, minimum height=20pt, align=left] (PlusTop) at (6,0.45){\Large{+}};
    \node[minimum width=20pt, minimum height=20pt, align=left] (PlusTop) at (5.55,0){\Large{+}};
    \draw [-] (ADD.135)--(ADD.315) ; \draw [-] (ADD.45)--(ADD.225) ;
    \draw [>=stealth,-, postaction={decorate}, thick, midway, above] ($(POD.east)+(0,1.2)$)--  (ADD.145) ;
    \draw [>=stealth,-, postaction={decorate}, thick, midway, above] ($(POD.east)+(0,0.4)$)--  (ADD.160) ;
    \draw [>=stealth,-, postaction={decorate}, thick, dashed, midway, above] ($(POD.east)+(0,0)$)-- node [midway,below left,yshift=-0.1cm, xshift=-0.3cm, rotate = 90] {$....$} (ADD.180) ;
    \draw [>=stealth,-, postaction={decorate}, thick, midway, above] ($(POD.east)+(0,-1.2)$)-- (ADD.215) ;
    
    \draw [-,thick] (ADD.35)--($(ADD.east)+(1,1.2)$) ;
    \draw [-,thick] (ADD.20)--($(ADD.east)+(1,0.4)$) ;
    \draw [-,thick,dashed] (ADD.east)--($(ADD.east)+(1,-0.6)$) ;
    \draw [-,thick] (ADD.-35)--($(ADD.east)+(1,-1.2)$) ;
    
    \draw [-,thick] ($(ADD.east)+(1,1.2)$)-- node [midway,above,align=right] {$\b{y}_{n+2}$}($(ADD.east)+(3,1.2)$) ;
    \draw [-,thick] ($(ADD.east)+(1,0.4)$)-- node [midway,above,align=right] {$\b{y}_{n+3}$}($(ADD.east)+(2.75,0.4)$) ;
    \draw [-,thick] ($(ADD.east)+(1,0.4)$)-- node [midway,below, yshift = -0.25cm,xshift = -0.25cm,align=right, rotate = 90] {$....$}($(ADD.east)+(2.75,0.4)$) ;
    \draw [-,thick,dashed] ($(ADD.east)+(1,-0.6)$)-- node [midway,above,align=right] {$\b{y}_{n+j}$}($(ADD.east)+(2.5,-0.6)$) ;
    \draw [-,thick] ($(ADD.east)+(1,-1.2)$)-- node [midway,above,align=right] {$\b{y}_{n+j+1}$}($(ADD.east)+(2.25,-1.2)$) ;
    
    \draw [-,thick]($(ADD.east)+(3,1.2)$) -- ++(0,-4) |- ($(ADD.east)+(3,-2.8)$)-|($(CNN.west)+(-2.5,0.4)$);
    \draw [-,thick]($(ADD.east)+(2.75,0.4)$) -- ++(0,-2.75) |- ($(ADD.east)+(2.75,-2.35)$)-|($(CNN.west)+(-2.25,0)$);
    \draw [-,thick,dashed]($(ADD.east)+(2.5,-0.6)$) -- ++(0,-1.4) |- ($(ADD.east)+(2.5,-2)$)-|($(CNN.west)+(-2,-1.2)$);
    
    \node[draw, thick, minimum width=20pt, minimum height=20pt, align=left] (Mean) at (6,1.5){\Large{$\b{\overline{y}}$}};
    \draw [>=stealth,-, postaction={decorate}, thick, midway] (Mean.south)-- (ADD.north) ;
   	\end{tikzpicture}
\caption{Hybrid Data-Driven Deep Learning Technique: Block diagram for the iterative prediction of the flow fields via deep learning CNN and POD-based low-dimensional approximation. The prediction starts from $\b{y}_{n+1}$ and loops until the desired number of time steps are predicted. The diagram illustrates the prediction of `$j$' number of flow fields from $\b{y}_{n+1}$ to $\b{y}_{n+j+1}$.}
\label{Fig:PredLoop}
\end{figure*}

After establishing the above steps, we can utilize the mean, the POD modes and the trained CNN to iteratively predict the future time steps starting from $\b{y}_{n+1}$. An illustration of the process is shown in figure \ref{Fig:PredLoop}. This technique is capable of predicting any number of time steps autonomously with the feedback loop provided that an adequate amount of high-fidelity data is used to determine the mean and the POD modes, and to train the CNN properly such that the prediction error is within the acceptable range for the predicted time series. 
We further elaborate on the integration of POD-based low-dimensional approximation with the CNN process.

\subsection{Convolutional Neural Network}
\label{CNN}
The convolutional neural network is designed to map the output and input taking the spatial features of the input into account. First, the CNN is trained using a known dataset and then used for predictions. The training phase consists of an iterative predictor (feed-forward) process and a corrector (back-propagation) process. Here, we briefly describe the components of the CNN in the context of flow field time series prediction. Refer to \cite{miyanawala2017efficient} for further elaboration on the use of CNN for fluid mechanics and \cite{goodfellow2016deep} for CNNs in general.
The training phase of the CNN involves the input function, the feed-forward process, and the back-propagation process.

Determination of the proper input and output fields is critical for the design of the CNN. For example, it is possible to map the flow fields $\{\b{y}_1,\b{y}_2,...,\b{y}_{n-1}\}$ directly to the flow fields of the adjacent time step $\{\b{y}_2,\b{y}_3,...,\b{y}_{n}\}$. This $m\times n-1$ to $m\times n-1$ mapping will require an extremely complex CNN with many layers. The training process will be computationally expensive and time-consuming and the network is prone to overfitting and large errors. Using the POD-based dimensionality reduction, we can first reduce the order of the output to $k \times n-1$ such that only the POD time coefficients $\{\b{A}_2, \b{A}_3,...,\b{A}_n\}$ are mapped to the flow field. The mean and the POD modes are stored to be used during the time series prediction of a dynamical system. Moreover, the CNN does not require the refinements and connectivities of the underlying mesh employed in a full-order solver. In the present work, a coarse 2D and uniformly ordered point-cloud with interpolated fluid field values are used as the input function. We consider a triangulation-based linear interpolation to convert the $\b{Y} \in \numberset{R}^{m \times n}$ dataset to the tensor $\b{Y}^* = \{\b{y}^*_1,\b{y}^*_2,...,\b{y}^*_{n}\} \in \numberset{R}^{p \times q \times n}$ where $p, q$ are the spatial dimensions of the 2D point-cloud.
It is worth mentioning that the size of $\b{y}^*_i$ is lower than a typical 2D mesh for a full-order simulation and is uniform regardless of the original mesh size. 
Once the input and output are determined, the CNN iteratively applies the feed-forward (predictor) and back-propagation (corrector) processes until the required convergence is achieved.

The feed-forward process applies the operations: convolution, rectification and down-sampling in sequence, first on the input matrix, then on the output obtained from the above operations. The convolution is a discrete operation which outputs a 3D tensor $\b{z}^c_i$ for each input $\b{y}^*_i$
\cite{goodfellow2016deep}:
\begin{equation}
\begin{split}
\label{Eq:Convolution}
\b{z}^c_i = \{{z}_{i\alpha\beta j}^c\} = \b{y}^*_i \star \b{K} = \left \{ \sum_{b=1}^{q}\sum_{a=1}^{p} y^*_{iab} K_{(\alpha-a+1)(\beta-b+1)j} \right \}, \\ j=1,2,...,n_k,  
\end{split}
\end{equation}
where $\b{K}$ is the adaptive time independent kernel tensor, $n_k$ is the number of convolution kernels and typically $\b{z}^c_i \in \numberset{R}^{p \times q \times n_k}$. The tensor changes slightly if the convolution stride $>1$ and any padding used, both of which are not incorporated in this paper. Further elaborations on the effect of non-unit stride and zero-padding can be found in 
\cite{goodfellow2016deep}.
Since convolution is a linear operation, the rectification process is incorporated to introduce nonlinearity to the training: $\b{z}^r_i = \max(\b{0},\b{z}^c_i)$. In a typical CNN, the rectification is followed by a down-sampling operation. However, in our study, we observe that it deteriorates the flow dynamic results. 
Usually, the convolution and rectification operations are repeated sequentially with kernel tensors with different sizing. Then the final tensor is stacked to get a vector $\b{z}_i \in \numberset{R}^{s}$, where $s$ is the stacked size of the final vector. This vector is then mapped to the output by simple weights:
\begin{equation}
\hat{\b{A}}_i = \b{w}\b{z}_i,
\end{equation}
where $\b{w} \in \numberset{R}^{k \times s}$ is the adjustable time-independent weight matrix and $\hat{\b{A}}_i$ is the predicted output of the CNN. This mapping is also known as a fully-connected layer. A CNN may contain more than one such fully-connected layers which are not discussed herein.  During the training, we calculate the error between the predicted ($\hat{\b{A}}_i$) and true ($\b{A}_i$) outputs and iteratively adjust the kernel tensors ($\b{K}$) and weight matrices ($\b{w}$) using the back-propagation technique.

The training phase of the CNN is an iterative process, which continuously minimizes the 
error between the predicted and actual outputs. 
The process compares the output of the feed-forward pass with the full-order result 
and corrects the kernels and weights to minimize the error. For the ease of explanation, let us denote these adjustable elements as $\bs{\mathcal{W}}$.
We define a cost 
function $\mathcal{C}$ to measure the discrepancy between the feed-forward 
prediction and full-order POD coefficients:
\begin{equation}
\b{E}_i = \mathcal{C}(\hat{\b{A}}_i,\b{A}_i).
\end{equation}
In this study, the cost function $\mathcal{C}$ is the root mean square error $L_2$ function. 
Now the target is to update the weight set $\bs{\mathcal{W}}$ to minimize the error $\b{E}_i$ using a gradient descent back-propagation method. In what follows is a brief description of the back-propagation.

For simplicity, let us denote the output of the $l^{th}$ layer (sub-routine) of the 
feed-forward process as $\b{z}^l_i$, which can be related to the previous layer 
output $\b{z}^{(l-1)_i}$ and the weights of the $l^{th}$ layer $\bs{\mathcal{W}}^l$ as follows:
\begin{equation}
\b{z}^l_i = \b{G}(\bs{\mathcal{W}}^l,\b{z}^{l-1}_i),
\end{equation}
where $\b{G}$ represent a single pass of convolution, rectification and down-sampling (if any) or a fully connected layer. 
Note that, for an $N$-layered CNN, $\b{z}^0_i=\b{y}^*_i$ and $\b{z}^{N}_i=\hat{\b{A}}_i$. 
The back-propagation process starts at the predicted value where the error gradient 
of the fully connected layer can be determined first. 
When the error gradient of the $l^{th}$ layer: $\frac{\partial \b{E}_i}{\partial \b{z}^l_i}$ is known, 
we can estimate the error gradients using the chain rule:
\begin{equation}
\begin{array}{cc}
\displaystyle \frac{\partial \b{E}_i}{\partial \bs{\mathcal{W}}^l} = \frac{\partial \b{G}}{\partial \bs{\mathcal{W}}}\left(\bs{\mathcal{W}}^l,\b{z}^{l-1}_i\right)\frac{\partial \b{E}_i}{\partial \b{z}^l_i},\\[8pt]
\displaystyle
\frac{\partial \b{E}_i}{\partial \b{z}^{l-1}_i} = \frac{\partial \b{G}}{\partial \b{z}}\left(\bs{\mathcal{W}}^l,\b{z}^{l-1}_i\right)\frac{\partial \b{E}_i}{\partial \b{z}^l_i},
\end{array}
\end{equation}
where $\frac{\partial \b{G}}{\partial \bs{\mathcal{W}}}$ and $\frac{\partial \b{G}}{\partial \b{z}}$ are the Jacobians of $\b{G}$ relative to $\bs{\mathcal{W}}$ and $\b{z}$, and are evaluated at the $l^{th}$ layer. 
After the evaluation of the error gradients  $\frac{\partial \b{E}_i}{\partial \bs{\mathcal{W}}}$, we employ the stochastic gradient descent method with momentum (SGDM) \cite{rumelhart1988learning} to adjust the parameter set for the $T^{th}$ iteration:
\begin{equation}
\bs{\mathcal{W}}_T = \bs{\mathcal{W}}_{T-1}-\gamma \frac{1}{S} \sum_{p=1}^S \frac{\partial \b{E}_i}{\partial \bs{\mathcal{W}}}+\psi (\bs{\mathcal{W}}_{T-1}-\bs{\mathcal{W}}_{T-2})
\label{Eq:Sgdm}
\end{equation}
where $\gamma>0$ is the learning rate, 
$\psi \in [0,1]$ is called the momentum and it is the hyper-parameter which 
determines the contribution from the previous gradient correction, and the parameter $S$ denotes the stochastic sample (mini-batch) size.
The gradient in the SGDM is an expectation with a provable convergence, which 
may be approximately estimated using a small set of samples \cite{lecun2012efficient}.
%
In the next section, we present the effectiveness of the POD-CNN technique for the predictions of flow fields and the wake dynamics behind a freely vibrating square cylinder in cross-flow.

\section{Results and discussion}
\subsection{Problem set-up}
\begin{figure}
\centering
\begin{subfigure}{0.5\textwidth}
\centering
\includegraphics[trim={0cm 0cm 0cm 0cm},clip,scale=0.375]{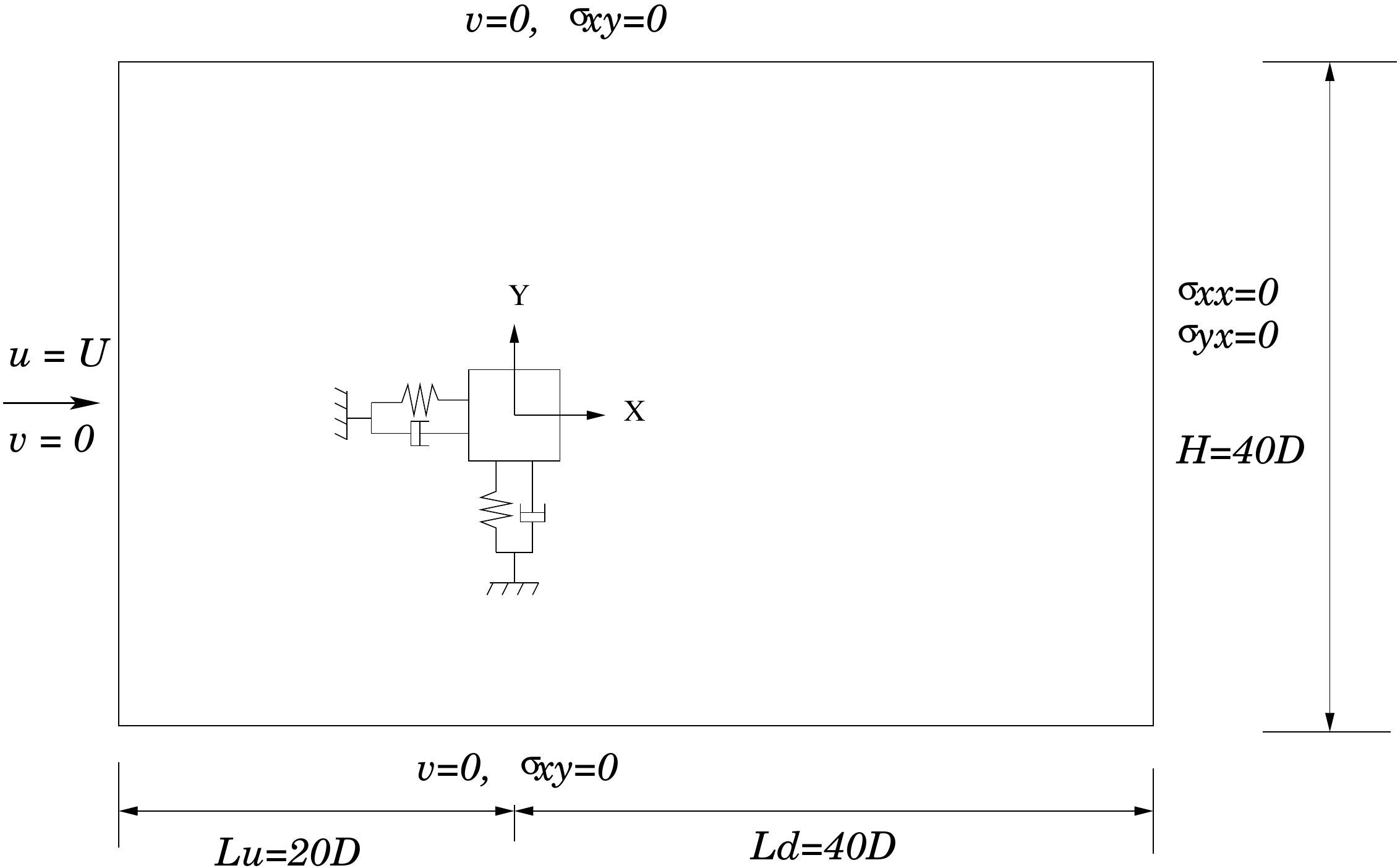}
\centering
\caption{}
\end{subfigure}
\begin{subfigure}{0.5\textwidth}
\centering
\includegraphics[trim={0cm 0cm 0cm 0cm},clip,scale=0.425]{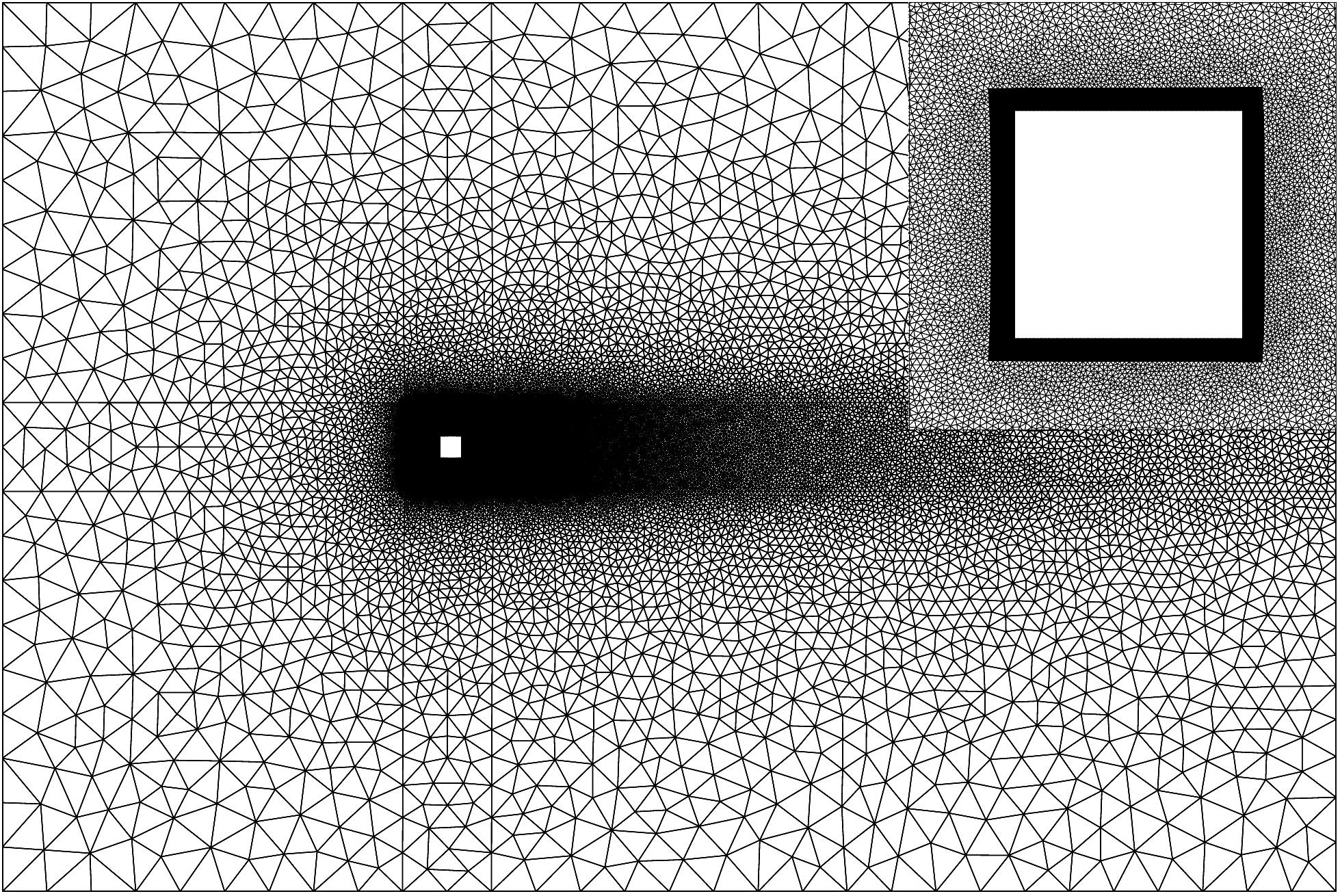}
\centering
\caption{}
\end{subfigure}
  \caption{(a) Schematic of the fluid domain, bluff body motion and boundary conditions. (b) Unstructured mesh used in finite element solver The top right inset shows the near cylinder mesh.}
  \label{fig:Schematic}
\end{figure}

In this work, for the first time, we apply a combined POD and CNN to determine the dynamic flow fields. We consider a prototypical fluid-structure interaction (FSI) set-up of an elastically-mounted cylinder to extract the unsteady flow field and to predict the dynamical data. 
In this coupled wake-cylinder problem, the low-dimensional flow structures such as the vortex street, the shear layer and the near-wake can be interpreted as an embedded feature into the high-dimensional snapshots.
The idea for the POD-CNN model is to extract these low-dimensional features and to predict the dynamic flow fields via the learned low-dimensional model.
In particular, we are interested to predict the time-series of the pressure field and the pressure force on a generic fluid-structure interaction problem.  
While the proposed technique can be generalized for any fluid-structure system involving the interaction
dynamics of flexible structures with an unsteady wake-vortex system, we consider a freely vibrating square cylinder in cross-flow, identical to \cite{miyanawala2018decomposition}.
The cylinder is free to oscillate 
in the streamwise ($X$) and the transverse ($Y$) directions. Figure \ref{fig:Schematic}(a) summarizes the fluid domain limits, degrees of freedom of the bluff body motion and the boundary conditions. Figure \ref{fig:Schematic}(b) is the unstructured mesh used in the full-order high-dimensional solver.
%

The wake-body synchronization is strongly influenced by 
the four key non-dimensional parameters, 
namely mass-ratio $\left(m^*=\frac{M}{m_f}\right)$, 
Reynolds number $\left(Re = \frac{\rho^\mathrm{f} U_{\infty} D}{\mu^\mathrm{f}}\right)$, reduced velocity $\left(U_r = \frac{U_{\infty}}{f_{n}D} \right)$,
and critical damping ratio $(\zeta=\frac{C}{2\sqrt{K M}})$
where $\rho^\mathrm{f}$ denotes the fluid density, $\mu^\mathrm{f}$ is the dynamic viscosity, $M$ is the mass per unit length of the body, $C$ and $K$ are the damping and stiffness 
coefficients, respectively for an equivalent spring-mass-damper system of a vibrating structure, 
$U_{\infty}$ and $D$ denote the free-stream speed and the diameter of the cylinder, respectively. 
The natural frequency of the body 
is given by $f_n=(1/2\pi) \sqrt{K/M}$ and  the mass of displaced fluid by the structure is $m_f = \rho^\mathrm{f} D^2 L_c$ for a square cross-section, 
and $L_c$ denotes the span of the cylinder. 
%
We particularly select these parameters to represent an FSI system with a low mass and damping ratios and undergoing synchronized wake-body motion. To obtain the high-dimensional approximation of the coupled wake-body interaction using the incompressible Navier-Stokes equations and the
rigid body dynamics, we employ the stabilized finite element method and a strongly coupled fluid-structure coupling scheme. Further details of the numerical methodology, the problem set-up and the discretization parameters can be found in \cite{miyanawala2018decomposition}.

We perform full-order simulations to extract an adequate number of pressure field snapshots. Using the flow field data, we first calculate the mean field, the POD basis vector and the POD time coefficients. Subsequently, we feed the pressure snapshots and POD coefficients as input-output pairs to the CNN. 
This allows to obtain the time series of the pressure and velocity fields within the error threshold. Using the pressure field, we evaluate the time history of the pressure-induced drag ($C_{Dp}$) and lift ($C_{Lp}$) forces given by:
\begin{equation}
\begin{split}
C_{Dp} = \frac{1}{\frac{1}{2}\rho^\mathrm{f}U_{\infty}^2D} 
\int_{\Gamma} (\bs{\sigma}_p \cdot \b{n}) \cdot \b{n}_x \: d\Gamma, \\ 
C_{Lp} = \frac{1}{\frac{1}{2}\rho^\mathrm{f}U_{\infty}^2D} 
\int_{\Gamma} (\bs{\sigma}_p \cdot \b{n}) \cdot \b{n}_y \: d\Gamma,
\end{split}
\end{equation}
where $\Gamma$ denotes the fluid-solid interface, $\bs{\sigma}_p = -p\b{I}$ is the pressure contributed stress component, $\b{n}_x$ and $\b{n}_y$ are the Cartesian components of the unit normal vector $\b{n}$. 
We further assess the POD-CNN based predictions with the full-order results in Section \ref{Results}. 

\subsection{Tuning of the neural network}
\label{HyperP}
\pgfplotsset{every tick label/.append style={font=\Large}}
\pgfplotsset{every tick label/.append style={scale=1}}
\pgfplotsset{
  log x ticks with fixed point/.style={
      xticklabel={
        \pgfkeys{/pgf/fpu=true}
        \pgfmathparse{exp(\tick)}%
        \pgfmathprintnumber[fixed relative, precision=3]{\pgfmathresult}
        \pgfkeys{/pgf/fpu=false}
      }
  },
  log y ticks with fixed point/.style={
      yticklabel={
        \pgfkeys{/pgf/fpu=true}
        \pgfmathparse{exp(\tick)}%
        \pgfmathprintnumber[fixed relative, precision=3]{\pgfmathresult}
        \pgfkeys{/pgf/fpu=false}
      }
  }
}
\begin{figure*}
\centering
\begin{subfigure}{0.5\textwidth}
\centering
\begin{tikzpicture}[trim axis left, 
trim axis right,
scale=0.85, 
baseline]
\begin{semilogyaxis}[
    xlabel={\Large Kernel size},
    ylabel={\Large $\max(MSE)$},
    xmin=1.8, xmax=8.2,
    xtick={2,4,6,8},
    xticklabels={$2 \times 2$,$4 \times 4$,$6 \times 6$,$8 \times 8$},
    width =9cm,
    height = 7cm,
    legend pos=north west,
    legend style={draw=none},
    legend columns = 2,
    legend style={at={(0.9,1.15)}, anchor=west},
]
 
\addplot[
    color=blue,
    dashed,
    mark=square,
    mark options = solid,
    very thick,
    mark size = 4,
    ]
    coordinates {
(2,0.0008933)(3,4.7224e-05)(4,7.2256e-05)(5,0.0004335)(6,0.0001249)(8,0.008102)
    };
    \addlegendentry{\Large POD-CNN}
    
    \addplot[
    color=red,
    solid,
    mark=o,
    very thick,
    mark size = 4,
    ]
    coordinates {
(2,2.0429e-05)(3,1.3450e-05)(4,1.5611e-05)(5,1.5925e-05)(6,7.6099e-06)(8,2.0415e-05)
    };
    \addlegendentry{\Large FOM}
\end{semilogyaxis}
\end{tikzpicture}
\caption{}
\end{subfigure}~
\begin{subfigure}{0.5\linewidth}
\vspace{0.75cm}
\centering
\begin{tikzpicture}[trim axis left, 
trim axis right,
scale=0.85, 
baseline]
\begin{semilogyaxis}[
    xlabel={\Large Number of kernels},
    ylabel={\Large $\max(MSE)$},
    xmin=9.8, xmax=100.2,
    xtick={10,20,30,40,50,60,70,80,90,100},
    width =9cm,
    height = 7cm,
    legend pos=north east,
    legend style={draw=none},
]
 
\addplot[
    color=blue,
    dashed,
    mark=square,
    mark options = solid,
    very thick,
    mark size = 4,
    ]
    coordinates {
(10,0.00014694)(20,0.00071734)(30,0.19602)(40,0.0002597)(50,0.00067742)(60,4.7224e-5)(70,0.00031306)(80,2.5570e-5)(90,5.3617e-5)(100,4.7086e-5)
    };
    
    \addplot[
    color=red,
    solid,
    mark=o,
    very thick,
    mark size = 4,
    ]
    coordinates {
(10,1.2600e-5)(20,2.0118e-5)(30,1.8672e-5)(40,1.5859e-5)(50,1.2795e-5)(60,1.2711e-5)(70,1.6548e-5)(80,1.3484e-5)(90,1.5741e-5)(100,1.8380e-5)
    };
\end{semilogyaxis}
\end{tikzpicture}
\caption{}
\end{subfigure}
\begin{subfigure}{0.49\linewidth}
\centering
\begin{tikzpicture}[trim axis left, 
trim axis right,
scale=0.85, 
baseline]
\begin{semilogyaxis}[
    xlabel={\Large Mini-batch size},
    ylabel={\Large $\max(MSE)$},
    xmin=9.8, xmax=60.2,
    xtick={10,20,30,40,50,60},
    width =9cm,
    height = 7cm,
    legend pos=north east,
    legend style={draw=none},
    legend columns = 2,
]
 
\addplot[
    color=blue,
    dashed,
    mark=square,
    mark options = solid,
    very thick,
    mark size = 4,
    ]
    coordinates {
(10,0.033681)(20,0.035025)(30,2.5570e-5)(40,9.5570e-5)(50,0.0064693)(60,0.0015758)
    };
    
    \addplot[
    color=red,
    solid,
    mark=o,
    very thick,
    mark size = 4,
    ]
    coordinates {
(10,7.0795e-6)(20,9.2713e-6)(30,1.3484e-5)(40,1.8570e-5)(50,2.2139e-5)(60,3.7381e-5)
    };
\end{semilogyaxis}
\end{tikzpicture}
\caption{}
\end{subfigure}~
\begin{subfigure}{0.49\linewidth}
\centering
\begin{tikzpicture}[trim axis left, 
trim axis right,
scale=0.85, 
baseline]
\begin{loglogaxis}[
    xlabel={\Large Learning rate ($\gamma$)},
    ylabel={\Large $\max(MSE)$},
    xmin=0.0009, xmax=0.055,
    xtick={0.001,0.005,0.01,0.025,0.05},
    log x ticks with fixed point,
    width =9cm,
    height = 7cm,
    legend style={draw=none},
    legend columns = 2,
]
 
\addplot[
    color=blue,
    dashed,
    mark=square,
    mark options = solid,
    very thick,
    mark size = 4,
    ]
    coordinates {
(0.001,0.0033219)(0.0025,0.0011785)(0.005,2.5570e-5)(0.01,0.0010083)(0.025,0.035072)(0.05,0.034852)
    };
    
    \addplot[
    color=red,
    solid,
    mark=o,
    very thick,
    mark size = 4,
    ]
    coordinates {
(0.001,0.00014718)(0.0025,0.00011056)(0.005,2.3480e-5)(0.01,1.1942e-5)(0.025,0.00034973)(0.05,0.00034806)
    };
\end{loglogaxis}
\end{tikzpicture}
\caption{}
\end{subfigure}
\caption{Hyper-parameter optimization of CNN: Variation of maximum MSE after 300 predictions (a) CNN kernel size (60 kernels, mini-batch size = 30, $\gamma=0.005$), (b) number of kernels ($3\times3$ kernels, mini-batch size = 30, $\gamma=0.005$), (c) mini-batch size (80, $3\times3$ kernels, $\gamma=0.005$) and (d) learning rate (80, $3\times3$ kernels, mini-batch size = 30). The trained network is fed iteratively by the predicted pressure fields (POD-CNN) and separately by the FOM pressure fields for the baseline comparison purpose. }
\label{fig:HyperPOpt}
\end{figure*}
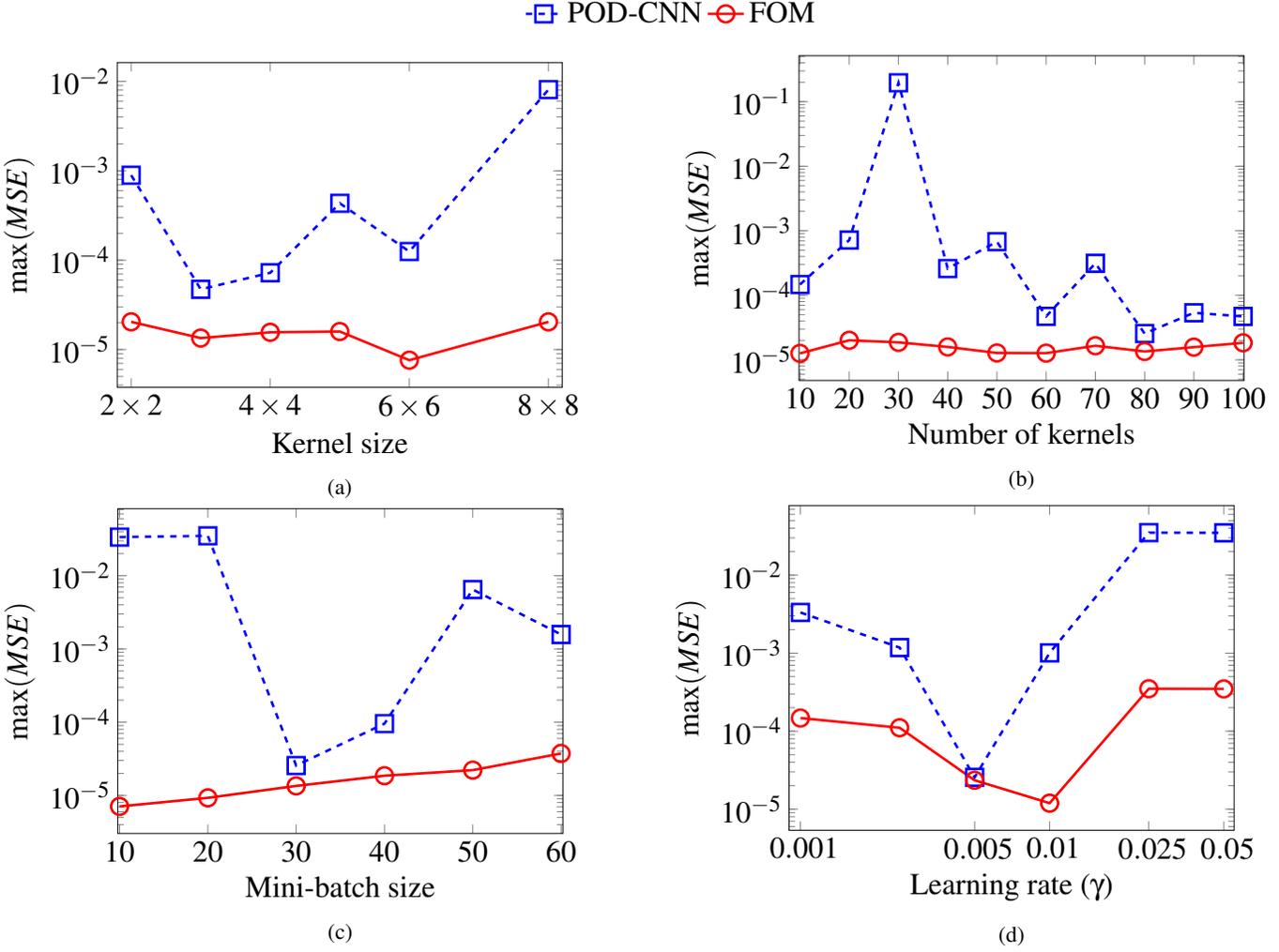

Extreme refining and overuse of the convolution layers may cause the CNN to overfit the training data set and make it incapable of predicting the flow fields other than the training dataset. However, the under-utilization of convolution will increase the error in the feature reconstruction and the prediction of long-term dynamics. Here, we present an empirical sensitivity study to establish the hyper-parameter values for the best performance of 
the CNN. Specifically, we address some of the general issues raised by \cite{kutz2017deep} for deep learning methods in fluid flows namely, the size of the kernels, the number of kernels per layer, mini-batch size and the learning rate to be used in the CNN. 
We carry out the FOM on a general laminar wake-body interaction case where the bluff body undergoes significant motion and exhibits statistically stationary flow dynamics. We consider a square cylinder FSI system with $Re=125, m^*=3.0, U_r=6.0$ and $\zeta=0.05$ and collect ($n=$)301 snapshots to obtain 300 future pressure fields via the POD-CNN procedure. To evaluate the trained model, the mean squared error (MSE) is calculated for each prediction which is given by:
\begin{equation}
MSE = \frac{1}{m} \sum_{i=1}^{m} \left(\hat{{P}}_i-{P}_i\right)^2,
\end{equation}
where $m$ is the mesh count, $\hat{{P}}_i$ and ${P}_i$ are the predicted and true normalized pressure of the $i^{th}$ node of the mesh, respectively. 
To demonstrate the POD-CNN learning model, we consider the maximum MSE of the 300 predicted pressure fields for the comparison purpose. Figure \ref{fig:HyperPOpt} summarizes the hyper-parameter optimization for our POD-CNN predictions. 
We only use one convolution layer, one rectification layer and one fully-connected layer for each POD mode. We assume that, in the convolution layer, 60 kernels have to be used with mini-batch size = 30 and learning rate ($\gamma$) = 0.005 and check the best kernel size. Subsequently, we optimize the number of kernels, the mini-batch size and the learning rate. Notably, we find that a network with a single convolution layer with 80 kernels of size $3\times3$, mini-batch size =30 and $\gamma=0.005$ gives the best performance. The pressure field prediction results of this CNN is presented in the next section.

\subsection{Pressure field reconstruction and prediction}
\label{Results}
We first use the CNN designed in Section \ref{HyperP} to extract the time-series of the pressure field of the laminar FSI system: a freely vibrating square cylinder in a uniform flow with $Re=125, m^*=3.0, U_r=6.0$ and $\zeta=0.05$. We obtain 301 pressure field snapshots (${\b{P}}_i,\: i=1,2,...,301$) using the FOM and calculate the mean ($\overline{\b{P}}$), the POD basis ($\bs{\mathcal{V}}$) and the POD coefficients (${\b{A}}_i,\: i=1,2,...,301$). As shown in figure \ref{fig:ModeEnergy}, the first 10 POD modes contain 99.99\% of the total mode energy. Hence we truncate $\bs{\mathcal{V}}$ and {$\b{A}$} only to contain the data of these 10 modes only. Refer to \cite{miyanawala2018decomposition} for further details on mode energy cascade.
After the truncation of the POD matrices, we feed the $(\b{P}_i,\b{A}_{i+1}),\: i=1,2,...,300$ pairs to train the CNN. The mean, the POD basis and the trained CNN are then employed to predict the pressure field time series: $\hat{\b{P}}_i, \: i=302,303,...,601$. The predictions are compared against the true pressure field series $\b{P}_i, \: i=302,303,...,601$ obtained by the FOM.
\pgfplotsset{every tick label/.append style={font=\Large}}
\pgfplotsset{every tick label/.append style={scale=1}}
\pgfplotsset{
  log x ticks with fixed point/.style={
      xticklabel={
        \pgfkeys{/pgf/fpu=true}
        \pgfmathparse{exp(\tick)}%
        \pgfmathprintnumber[fixed relative, precision=3]{\pgfmathresult}
        \pgfkeys{/pgf/fpu=false}
      }
  },
  log y ticks with fixed point/.style={
      yticklabel={
        \pgfkeys{/pgf/fpu=true}
        \pgfmathparse{exp(\tick)}%
        \pgfmathprintnumber[fixed relative, precision=3]{\pgfmathresult}
        \pgfkeys{/pgf/fpu=false}
      }
  }
}
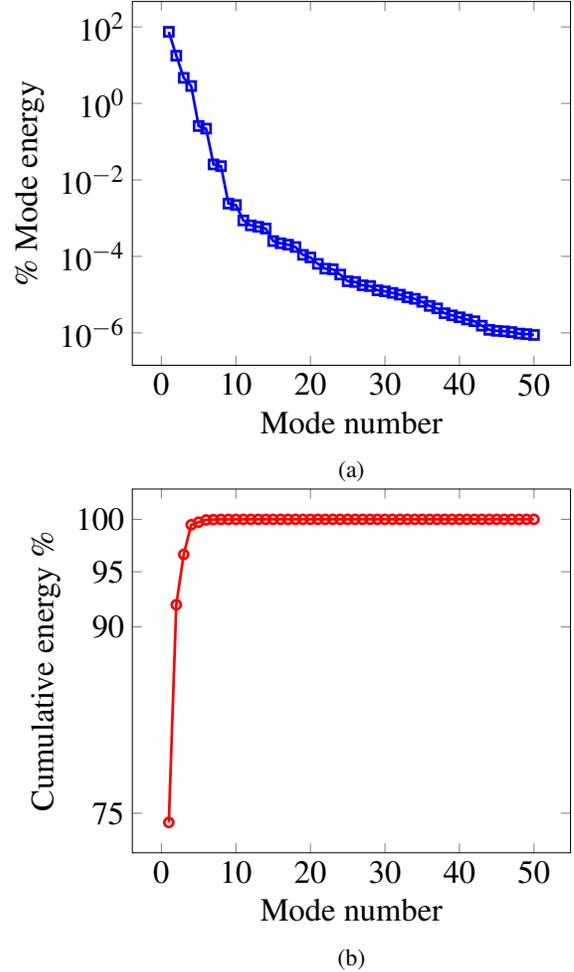
\begin{figure}
\centering
\begin{subfigure}{0.5\textwidth}
\centering
\begin{tikzpicture}[trim axis left, 
trim axis right,
scale=0.85, 
baseline]
\begin{semilogyaxis}[
    xlabel={\Large Mode number},
    ylabel={\Large $\%$ Mode energy},
]
 
\addplot [color=blue,
    solid,
    mark=square,
    mark options = solid,
    very thick,
    mark size = 2,] table [x = Mode, y index=1
    ] {data.txt};
\end{semilogyaxis}
\end{tikzpicture}
\caption{}
\end{subfigure}
\begin{subfigure}{0.5\textwidth}
\centering
\begin{tikzpicture}[trim axis left, 
trim axis right,
scale=0.85, 
baseline]
\begin{semilogyaxis}[
    xlabel={\Large Mode number},
    ylabel={\Large Cumulative energy $\%$},
     ytick={75, 90, 95, 100},
     yticklabels={75, 90, 95, 100},
]
 
\addplot [color=red,
    solid,
    mark=o,
    mark options = solid,
    very thick,
    mark size = 2,] table [x = Mode, y index=2
    ] {data.txt};
\end{semilogyaxis}
\end{tikzpicture}
\caption{}
\end{subfigure}
\caption{Mode energy cascade of the POD modes: (a) Percentage energy and (b) Cumulative energy of the first 50 modes out of the total 300 modes.}
\label{fig:ModeEnergy}
\end{figure}

\begin{figure}
\centering
\begin{subfigure}{0.5\textwidth}
\centering
\includegraphics[trim={0cm 0.25cm 0.25cm 1cm},clip,scale=0.275]{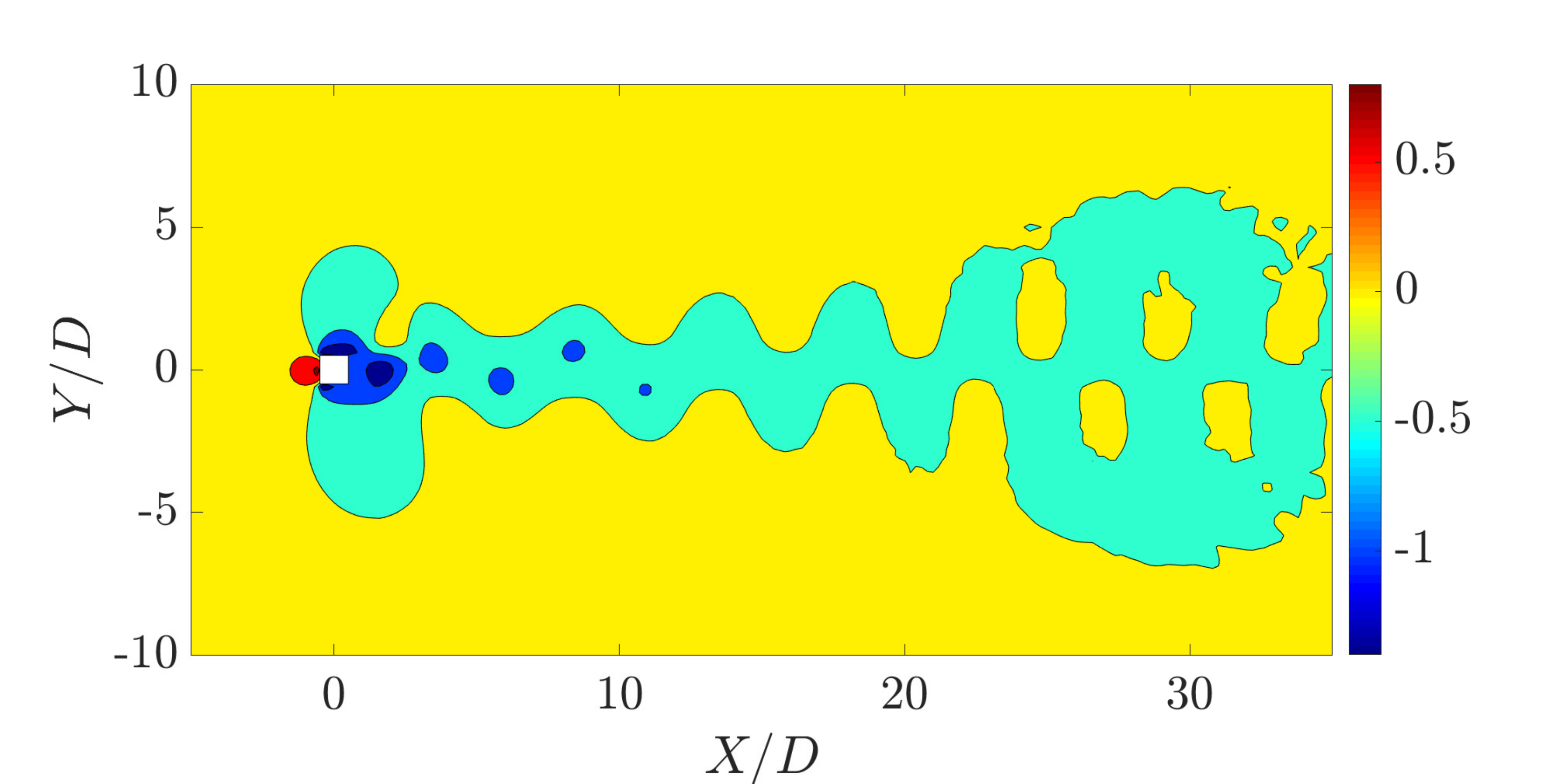}
\caption{}
\end{subfigure}
\begin{subfigure}{0.5\textwidth}
\centering
\includegraphics[trim={0cm 0.25cm 0.25cm 1cm},clip,scale=0.275]{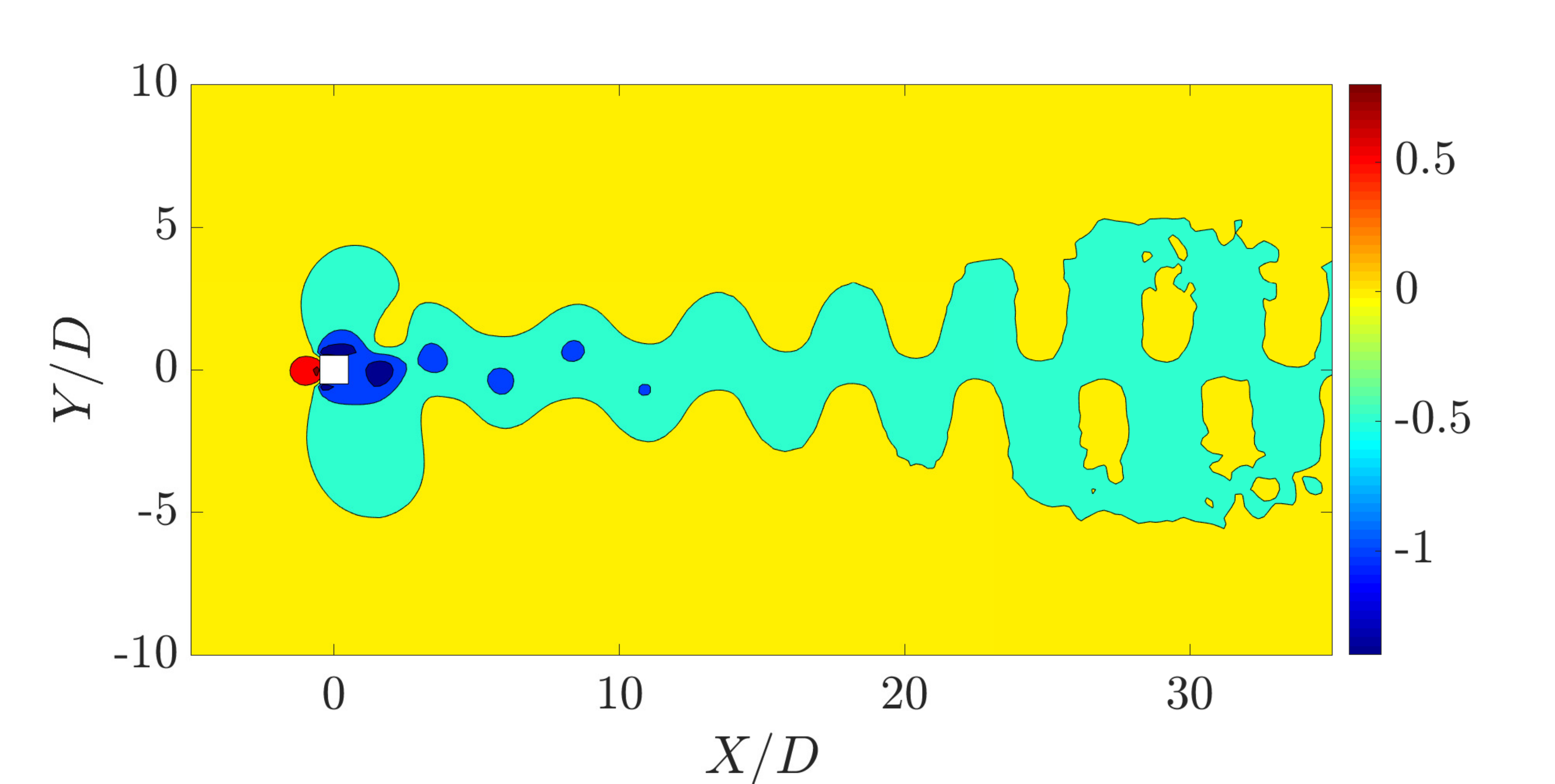}
\caption{}
\end{subfigure}
\begin{subfigure}{0.5\textwidth}
\centering
\includegraphics[trim={0cm 0.25cm 0.25cm 1cm},clip,scale=0.275]{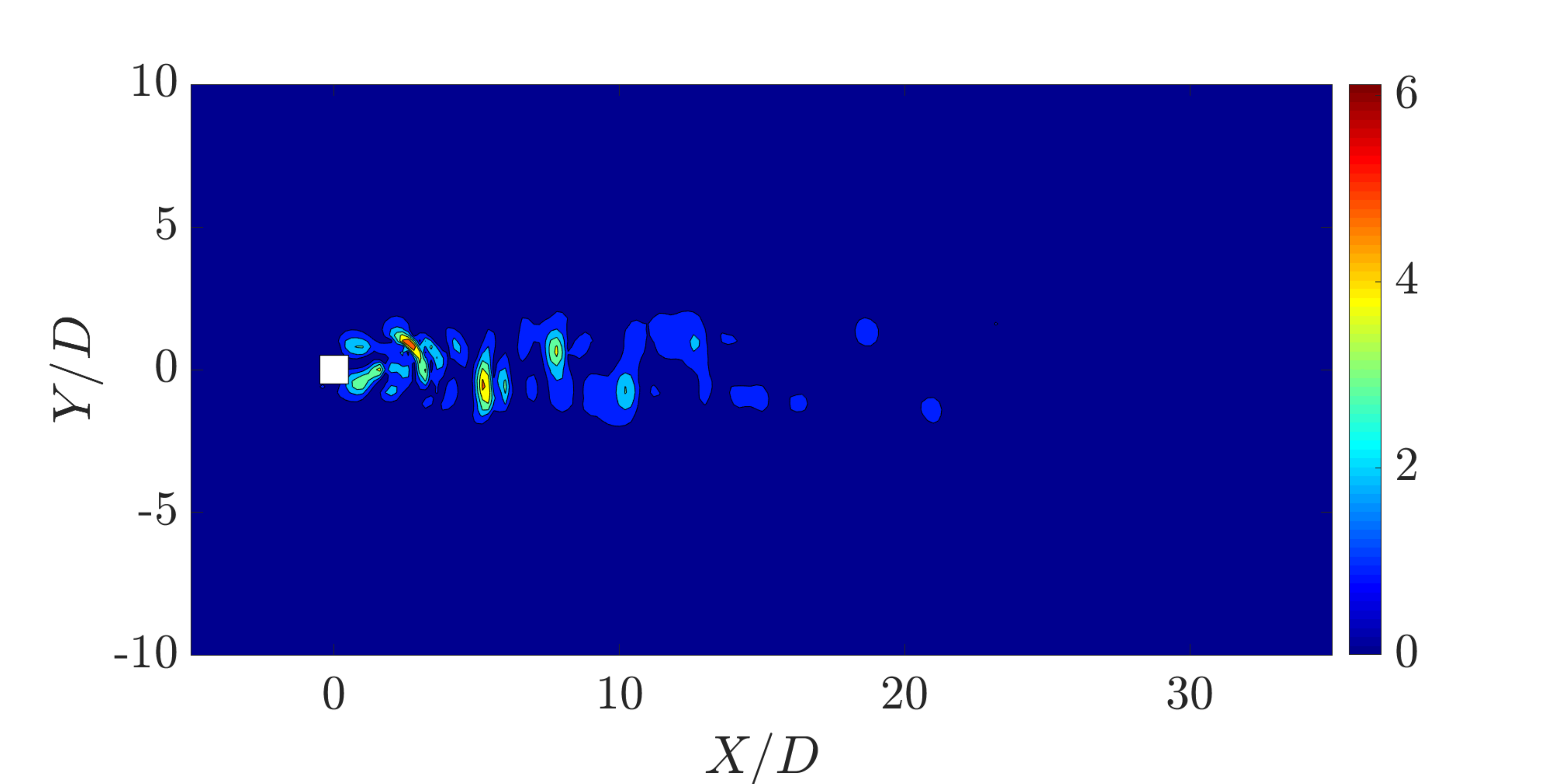}
\caption{}
\end{subfigure}
\begin{subfigure}{0.5\textwidth}
\centering
\includegraphics[trim={0.75cm 0cm 0.75cm 0.75cm},clip,scale=0.24]{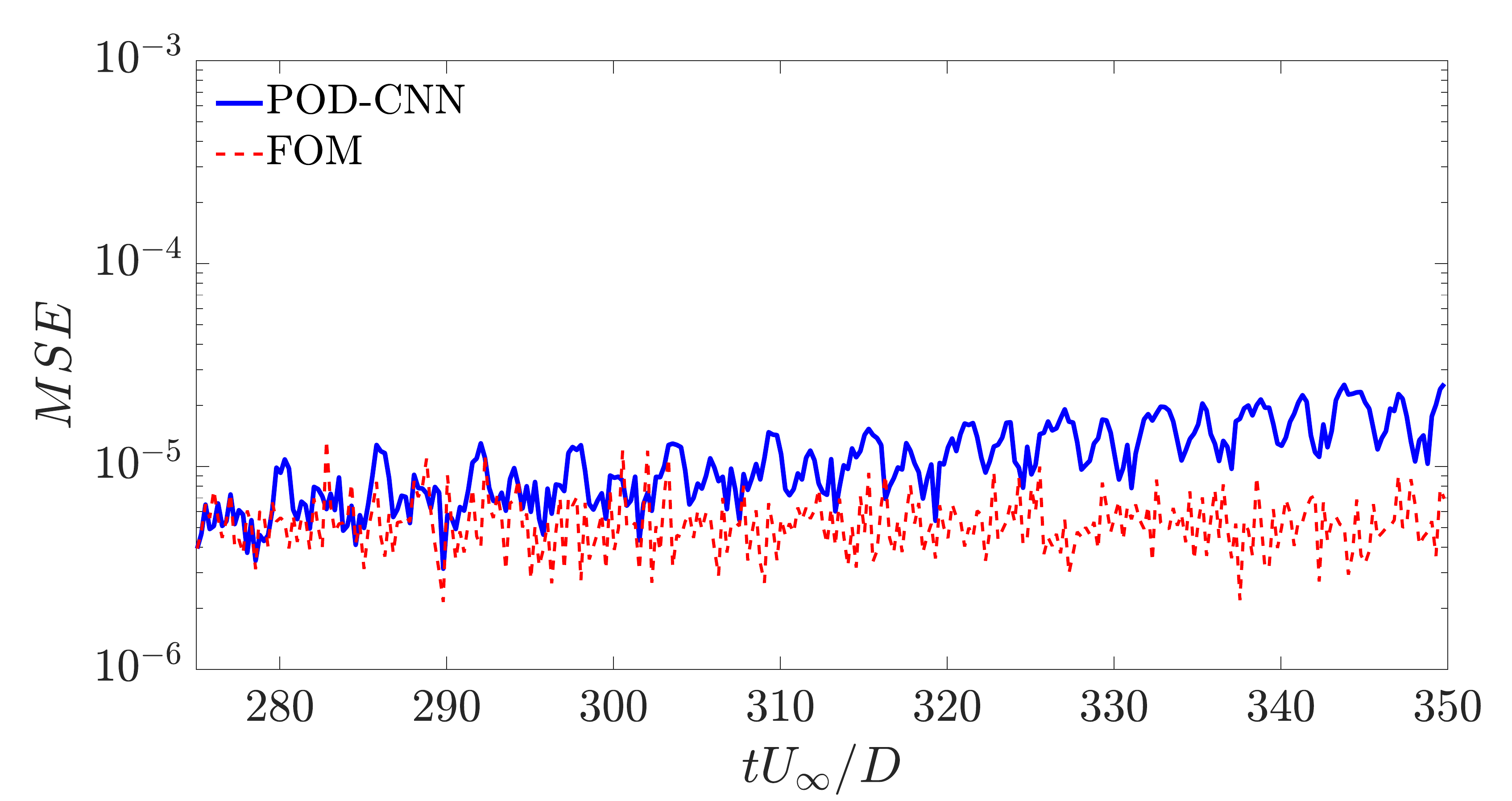}
\centering
\caption{}
\end{subfigure}
\caption{Accuracy of the POD-CNN for a vibrating cylinder at $Re=125, U_r=6.0, m^*=3.0$: (a) prediction, (b) original pressure fields, (c) normalized error at 300th time step, (d) MSE variation at first 300 predictions. The flow is from left to right.}
\label{fig:PredictionError}
\end{figure}
\begin{figure}
\centering
\begin{subfigure}{0.5\textwidth}
\centering
\includegraphics[trim={0cm 0cm 0.25cm 0.25cm},clip,scale=0.26]{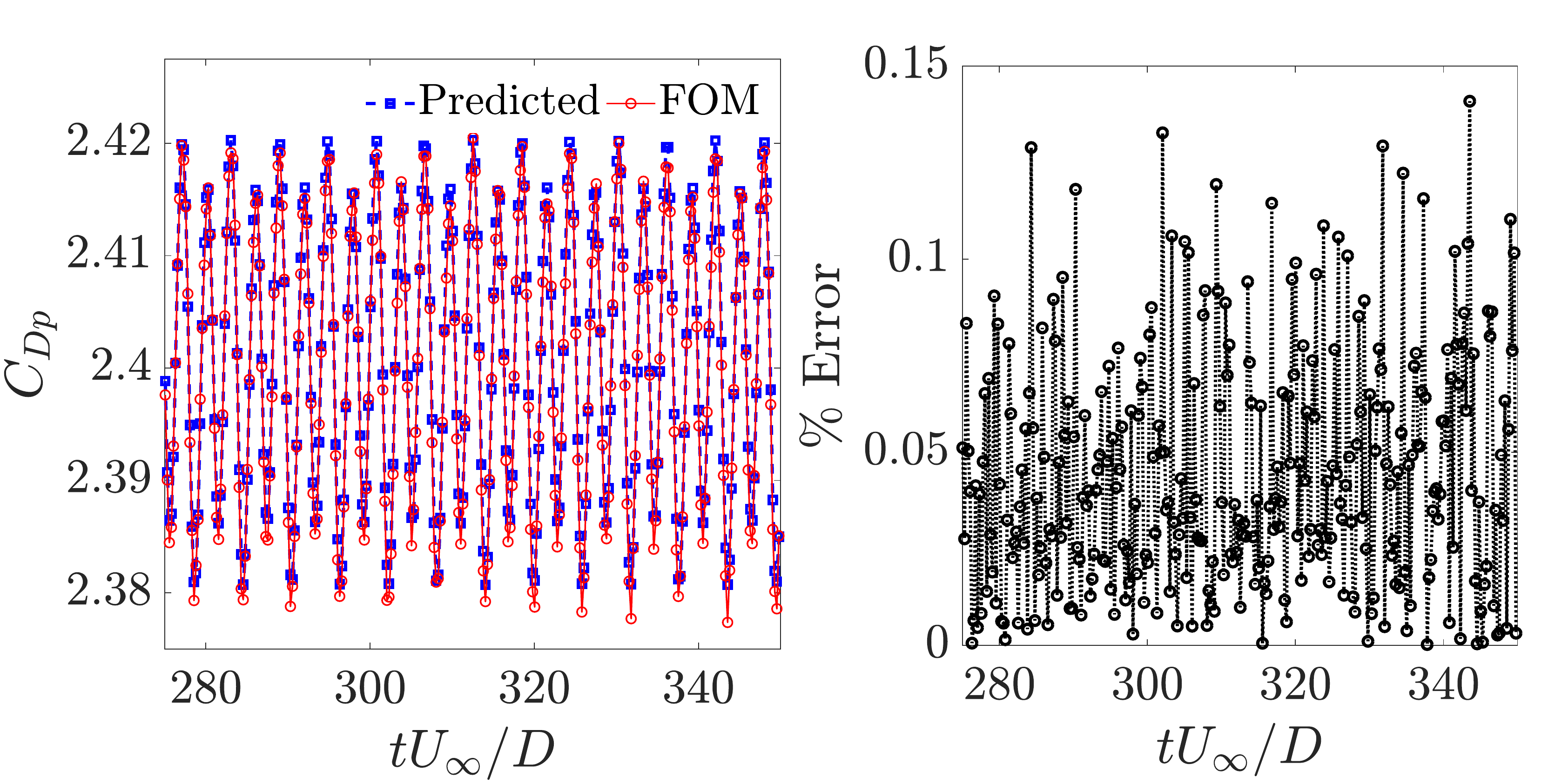}
\centering
\caption{}
\end{subfigure}
\begin{subfigure}{0.5\textwidth}
\centering
\includegraphics[trim={0cm 0cm 0.25cm 0.25cm},clip,scale=0.26]{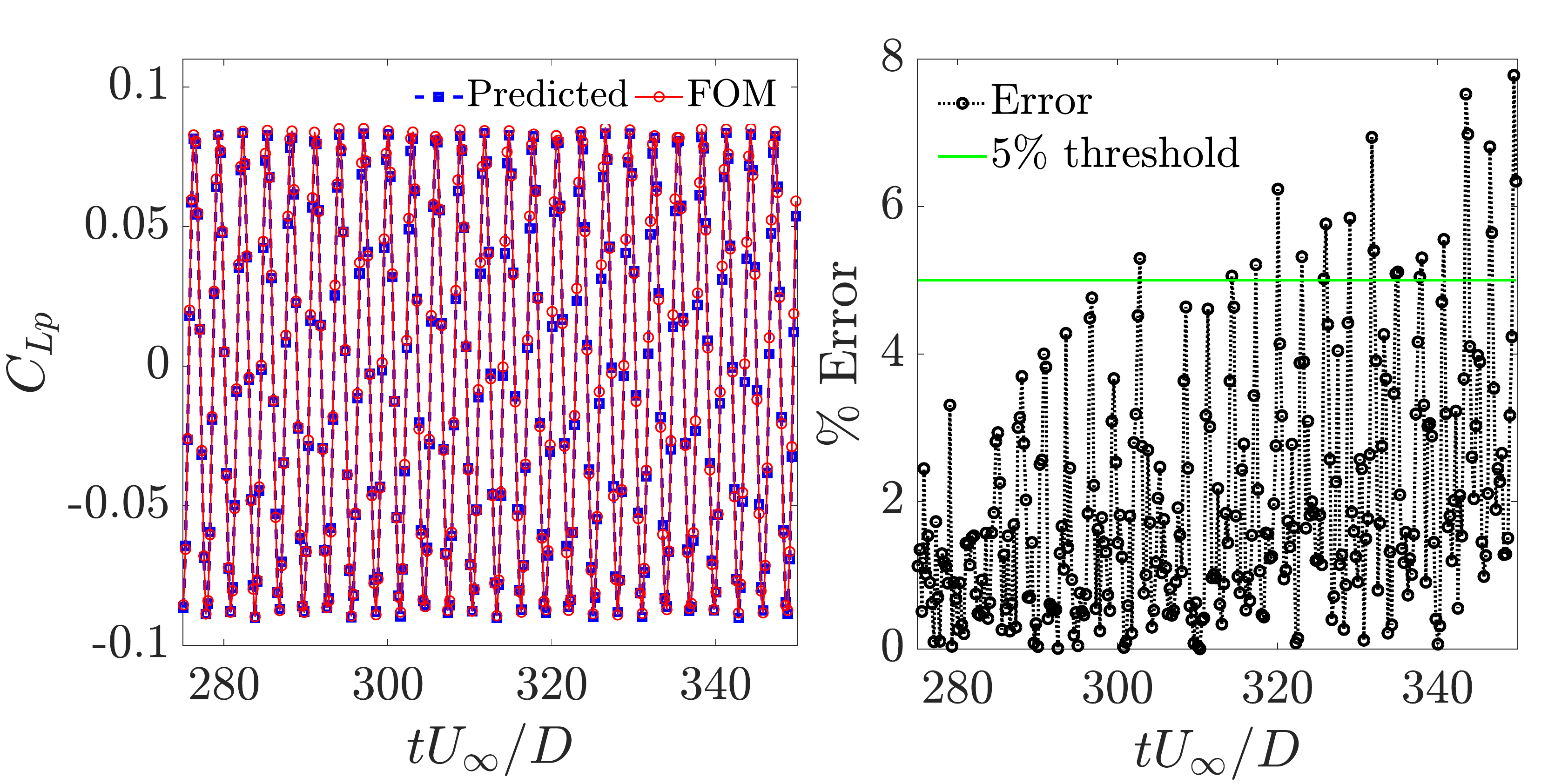}
\centering
\caption{}
\end{subfigure}
  \caption{Predicted (e) pressure drag, and (f) lift variation along with associated error.  Pressure values are normalized by $\frac{1}{2}\rho U_{\infty}^2$.}
  \label{fig:PredictionErrorP}
\end{figure}
Figure \ref{fig:PredictionError} illustrates the prediction quality of the POD-CNN technique. For the first time, we are able to extract the flow field without the use of a full-order simulation and yet maintain a remarkable accuracy. 
Figure \ref{fig:PredictionError}(a) and figure \ref{fig:PredictionError}(b) show the predicted and FOM pressure fields for the 300th time step ($\hat{\b{P}}_{601}$ and ${\b{P}}_{601}$). Figure \ref{fig:PredictionError}(c) displays the percentage error between these two fields, whereas the error is below 6.2\%. 
The proposed POD-CNN model maintains the mean-squared error of 300 flow fields below $3\times10^{-5}$ (figure \ref{fig:PredictionError}(d)) hence the POD-CNN model has learned to predict the unsteady flow fields over a broad range of time steps. For the comparison purpose, we feed the POD-CNN with the true pressure fields and evaluate the error of the predictions. The error begins to increase as a function of time, while the MSEs of the predictions are maintained between $[2\times10^{-6}, 2\times10^{-5}]$. 
The iterative prediction gives almost the same accuracy for the entire time range.

Furthermore, the streamwise (drag) and the cross-flow (lift) forces due to the pressure field are extracted for both POD-CNN and FOM simulations. We have a good match between the results (figures \ref{fig:PredictionErrorP} (a) and (b)). The pressure drag prediction has an accuracy of 99.8\%. 
Considering the 
lift force due to the pressure field, all the predictions have a 92\% accuracy and 279/300 (93\%) of the time steps are above the 95\% accuracy threshold.

\begin{table}[h]
\centering
\caption{Summary of computational resources used, speed-up and error between FOM and POD-CNN predictions. While FOM is performed on a multi-core workstation (HPC) and a single-core personal computer (PC) system, the POD-CNN computation is done on a PC. The elapsed times are estimated to obtain 300 time steps at $0.25D/U_{\infty}$ intervals. The accuracy values are given for the pressure field of the 300th time step predicted. 
}{\label{FOMvsCNN}}
\begin{tabular}{l|c|c|c}
                           & HPC & PC & POD-CNN \\ \hline
No. of Processors          & 24      & 1       & 1 \\
Processor (GHz)            & 2.60   & 2.60  & 2.60 \\
RAM (GB)                   & 256     & 16      & 16\\
Time elapsed (s)          & 798    & 19064    & 157\\
Speed-up (FOM-HPC)         &  &  & $5.08$ \\
Speed-up (FOM-PC)          &  &  & $121.4$ \\
Domain error (MSE)         &  &  & $\leq 3\times10^{-5}$ \\ 
Local accuracy             &  &  & $\geq$93.8\% \\\hline
\end{tabular}
\end{table}
Table \ref{FOMvsCNN} summarizes the performance of the POD-CNN technique. Full-order model is run on a multicore high-performance computer (HPC) and a single-core personal computer (PC) and the POD-CNN is performed on the same PC. While the parallel simulations utilize 24 processors and 256GB RAM for each case, the serial PC simulations are performed on a single processor with 16GB RAM. We run each simulation to extract 300 snapshots at $0.25D/U_{\infty}$ which encloses $\sim 12$ shedding cycles. Using the serial computation power of a PC, POD-CNN performs $\sim 5$ times faster in comparison to the FOM simulations on an HPC which has 24 times the processing power and 16 times the RAM. When the FOM and POD-CNN are performed in the same PC, the POD-CNN method is $\sim 120$ times faster than the FOM-PC. In summary, the POD-CNN method predicts the unsteady flow-fields for long time series with a domain mean squared error $\leq 3\times 10^{-5}$ and local accuracy $\geq 93.8\%$ using a small fraction of the computational resources as compared to its FOM counterpart.  

The present POD-CNN technique is capable of predicting the nonlinear unsteady flow fields for the same fine mesh used for full-order simulations while maintaining good qualitative and quantitative agreement. 
As a result, the physical quantities derived from the predicted flow field such as the pressure forces are found to be reasonably accurate from offshore engineering analysis and operational standpoints. 
Although a canonical laminar flow of a freely vibrating square is considered, the present POD-CNN learning technique does not make any assumptions with regard to geometry and boundary conditions hence forms a general framework for predicting future measurements of a large-scale offshore dynamical system.


\section{Concluding remarks}
\label{Conclu}
We have presented a novel deep learning model based on the POD and CNN approximations for unsteady wake flows and wake-body interaction. While the POD provides an optimal extraction of dominant flow features, the CNN enables nonlinear correlation of fluid motions to construct time-series of flow field using the trained neural networks. We successfully demonstrated the effectiveness of the proposed technique by predicting the time series of the unsteady pressure field of an FSI set-up of an elastically-mounted square cylinder.
We have tuned the hyper-parameters namely the number of kernels, the kernel size, the training batch size and the learning rate of the CNN to provide the most accurate reconstruction and prediction of the flow field. For the first time, using the low-dimensional approximation and the CNN-based learning, we are able to predict the flow patterns of a long time series using the same full-order mesh with reasonable accuracy. The quantities derived from the flow fields such as the pressure drag and lift were also accurately predicted.
%
The present low-dimensional learning technique can be implemented using only a fraction of computational resources to that of FOM and allows rapid exploration of flow pattern variations due to the design parameter variations (e.g., bluff body geometry and structural parameters). 
The POD-CNN technique presents a general framework for predicting any real world fluid flow scenario or design and it does not make any assumptions with regard to the problem specifications. 
Hence, the POD-CNN method can be developed to be the flow pattern prediction tool for real world scenarios such as large scale offshore structures which consist of turbulent three-dimensional FSI flows and complex bluff body geometries with multidegrees of freedom motion. The proposed data-driven deep learning framework can be used for efficient design and optimization, parametric fluid-structure analysis, real-time monitoring and control of offshore structures.

\section*{Acknowledgements}
The first author thanks the Ministry of Education, Singapore for the financial support. 
\bibliographystyle{asmems4}
\bibliography{CNNPODRef}
\end{document}